\documentstyle[12pt,A4,fleqn,epsfig]{article}
\begin{document}

 \hspace{10cm} DFUB 99/20 \par
\parskip -1pt
 \hspace{10cm} SLIM 99/1 \par

\begin{center}

{\large\bf SEARCH FOR  ``LIGHT'' MAGNETIC MONOPOLES }
\vspace{5mm}

The SLIM Collaboration
\end{center}

\vskip .2 cm \begin{center}
\nobreak\bigskip\nobreak
\pretolerance=10000

D.~Bakari$^{1,4}$,
S.~Cecchini$^{1,a}$, 
H.~Dekhissi$^{1,4}$, 
J.~Derkaoui$^{1,4}$, 
G.~Giacomelli$^{1}$,
M.~Giorgini$^{1}$, 
J.~McDonald$^{3}$, 
G.~Mandrioli$^{1}$, 
S.~Manzoor$^{1,5}$, 
A.~Margiotta$^{1}$, 
A.~Marzari-Chiesa$^{2}$, 
J.~Nogales$^{7}$,
M.~Ouchrif $^{1,4}$, 
L.~Patrizii$^{1}$, 
J.~Pinfold$^{3}$, 
V.~Popa$^{1,6}$, 
O.~Saavedra$^{2}$, 
P.~Serra$^{1}$, 
M.~Spurio$^{1}$, 
R.~Ticona$^{7}$, 
V.~Togo$^{1}$,
A.~Velarde$^{7}$,
E.~Vilela$^{1}$
and A.~Zanini$^{2}$

\vspace{0.6cm}

\footnotesize 

1. Dipar.to di Fisica dell'Universit\`a di Bologna and INFN,
 40127 Bologna, Italy \\
2. Dipar.ti di Fisica Sperimentale e Generale dell'Universit\`a di Torino and INFN,
 10125 Torino,  Italy \\
3. Centre for Subatomic Research, Univ. of Alberta, Edmonton, 
Alberta T6G 2N4, Canada\\
4. Faculty of Sciences, University Mohamed I, B.P. 424, Oujda, Morocco \\
5. RPD, PINSTECH, P.O. Nilore, Islamabad, Pakistan \\
6. Institute for Space Sciences, 76900 Bucharest, Romania\\
7. Laboratorio de Fisica Cosmica de Chacaltaya,UMSA, La Paz, Bolivia\\
$a$ Also Istituto TESRE/CNR, 40129 Bologna, Italy \\
\end{center}

\vskip 0.5 truecm
\begin{abstract}
We propose to implement one passive nuclear track detector array  
of $400$~ m$^2$ at the 
Chacaltaya High Altitude Laboratory (5230 m a.s.l.).
 The main purposes of the experiment
 concern the  searches 
for magnetic monopoles of relatively low masses at the  
 Parker bound level,  searches 
for low mass nuclearites, 
and searches for some supersymmetric 
dark matter candidates (Q-balls).\par
\end{abstract}

\vskip 1.0 truecm
 \noindent{\bf 1. Introduction}\par
\vskip 0.5 truecm

The search for magnetic monopoles (MMs) in the penetrating 
cosmic radiation remains one of the main items of non-accelerator particle 
astrophysics.\par

Grand Unified Theories (GUT) of electroweak and strong interactions 
predict the existence of superheavy magnetic monopoles with  
masses larger than $10^{16}$ GeV \cite{gg}. They would have been produced 
at the end of the GUT epoch, at a mass scale $\sim 10^{14}$ GeV and
 the cosmic time of $ \sim 10^{-34}$ s. 
Such monopoles cannot be produced 
with existing accelerators, nor 
 with any foreseen for the future. 
The MACRO experiment is well suited for their study and is providing the best 
experimental limits \cite{macro}.

Lower mass monopoles, proposed by many authors, require 
a phase transition in the early universe in which a semisimple gauge group 
yields a U(1)
factor at a lower energy scale [1, 3]. MMs with masses around
$10^{6} \div 10^{10}$ GeV have been proposed [3-6].
A MM is a topological point defect; an
undesirable large number of relatively light monopoles may be gotten rid of by means of higher
dimensional topological defects (strings, walls) \cite{huguet}.\par
One of the recent interests in relatively low mass MMs is connected also
with the possibility that relativistic MMs could be the source of the
highest energy cosmic rays, with energies larger than $10^{20}$ eV [3-5].
For monopoles one knows possible acceleration 
mechanisms: since the basic magnetic charge should be very large, relatively
light monopoles
can be accelerated to relativistic velocities and to energies of the order 
of $10^{20}$ GeV in one coherent domain of the galactic
 magnetic field, or in the intergalactic field, or  in many astrophysical 
sites, 
like in the magnetic field of Active Galactic Nuclei (AGN)
 and even of neutron stars. The next problem is how
these monopoles can interact in the upper atmosphere and yield 
electromagnetic 
showers. This is for istance possible if a monopole forms a bound state  with
 a proton (a dyonic system) which then may interact with a cross section 
typical of a relativistic hadron ($\sigma \geq 10^{-26}$ cm$^2$) [3-5]. 
Monopole masses of $10^{6}\div 10^{10}$ GeV could be consistent with a flux at
the 
Parker limit \cite{huguet,parker}.\par
We must also remember  
the possibility  that MMs could be multiply charged, $g=2g_D$, as in some SUSY 
theories, and $g=3g_D$, as in some superstring models 
($g_D=\hbar c/2e=68.5~e$ is the basic Dirac monopole charge) and that the 
basic charge could be $1/3~e$ \cite{bps}.
\par
We propose a search for relatively light MMs with an array of 400 m$^2$ of 
passive 
nuclear track detectors deployed at the high altitude Chacaltaya lab (5230 m 
high above sea level).
In $>4$ years of operation we should be able to reach a 
sensitivity at the level 
of the Parker bound, i.e.
$\sim 10^{-15}$ cm$^{-2}$ s$^{-1}$ sr$^{-1}$. 
  Byproducts of this MM  search are the searches 
for relatively light nuclearites  and Q-balls \cite{macron}.\par
We recall that nuclearites (strangelets, strange quark matter) are
nuggets of strange 
quark matter 
(aggregates of $u$, $d$, and $s$ quarks in equal proportions);
they could be the ground state of QCD and could be part of the cold dark
matter, and could have typical galactic velocities $\beta \sim 10^{-3}$
\cite{wit}. \par
Q-balls are supersymmetric coherent states of squarks, sleptons and Higgs 
fields, predicted by minimal 
supersymmetric generalizations of the Standard Model; they 
 could be copiously produced in the early universe. Relic Q-balls are 
also candidates for the cold dark matter \cite{qb}.\par
Since both nuclearites and charged Q-balls lose a large amount of energy
for $\beta~>~ 4~\times 10^{-5}$ they would be easily detectable 
with the proposed track-etch system. \par

An exposure at a high altitude laboratory would
allow to search for MMs of lower masses, 
higher magnetic charges and 
lower velocities \cite{derkaoui},
see Figs.~1 and 2.
The same  holds for lighter nuclearites and Q-balls.
For low mass nuclearites one would 
reach  a level of sensitivity more than one order 
of magnitude lower than any of the 
existing limits \cite{naka}.

Fig. 3 shows the accessible region in the plane (mass, $\beta$) for 
nuclearites, 
at MACRO depth, at ground level (under 1000 g cm$^{-2}$ of atmosphere), 
at the Chacaltaya altitude (540~g~cm$^{-2}$ of atmosphere) and at 20 km height
 assuming that the nuclearites have standard energy losses \cite{wit}. 
Lower mass nuclearites should be much more abundant than higher mass 
ones \cite{wilk3}.

The high altitude exposure would allow detection of the above mentioned 
particles even if they had strong interaction cross sections which could 
prevent  
them from reaching the earth surface.
From this point of view, it is important that the site be at the highest 
altitude. 
(On this point we have asked the opinions of several colleagues. In particular S. Glashow, 
J. Steinberger and A. De R\'{u}jula adviced us to search for the above
 mentioned objects at as high an altitude as possible.)
\par

Experimental data obtained at the highest altitude laboratories suggest
the existence of ``Centauro events" and other exotic events. 
It 
has also been suggested  that nuclearites with 
mass number of only few hundred could have 
nuclearite~-~air nuclei collisions at high altitudes in which  
the baryonic number of the nuclearite reduces by about the mass number of 
the target nucleus; this effect can be neglected for large mass nuclearites 
but it could 
seriously alter the energy loss of nuclearites with $A \sim 1000$ \cite
{centauro,wilk}. 
The
nuclearites  would decrease in $A$ in successive interactions with air nuclei 
until reaching some critical value, $A_{crit} \sim 320$, below which they 
disintegrate into nucleons. This mechanism could be tested at the highest
altitude stations.\par
 According to  
 ref. \cite{wilk2} the initial mass number 
that a nuclearite
should have at the entry in the atmosphere in order to reach the Chacaltaya 
altitude (540 g cm$^{-2}$) before disintegrating into normal baryons is  
$A \simeq 3000$, while to reach  
450 g cm$^{-2}$  altitude the initial mass should be about $A \simeq 1700$. 
Assuming that the abundance of 
nuclearites outside the atmosphere 
has the same $A$ dependence as the abundance of elements in the Universe, 
the flux for $A \sim 1700$ would be about 100 times larger than the flux 
of nuclearites with $A \sim 3000$ \cite{wilk3}.    \par
We hope to instrument a small area ($\sim$ 2 m$^{2}$) 
with many layers of thin CR39
and other detectors; this also applies to a very small detector at higher
altitudes.

\vskip 1.0 truecm
 \noindent{\bf 2. Experimental method}\par
\vskip 0.5 truecm

In order to reach 
the goal of a sensitivity at the level of the Parker bound 
\cite{parker}, one needs a surface detector of 
the least 400 m$^2$ and operation for at least four years. 
This detector also yields good limits on lighter nuclearites and Q-balls.

 In order to achieve 
the best redundancy and ``convincingness" the best detector should 
have redundant types of subdetectors, such as those presently
used by the MACRO experiment at Gran Sasso: i) liquid scintillators 
for wave form shape and time-of flight (ToF) informations, ii) tracking 
system to ensure  
single space track and single \lq\lq time track" (for slow monopoles); iii) 
passive 
nuclear track detectors for space track and restricted energy loss 
analyses \cite{macrom}. 
This solution would also yield byproducts on cosmic ray 
physics  \cite{rep}, but it would be 
 expensive and complicated and it is not needed for the limited purposes of 
this proposal.
\par

For the near future 
 the simplest possibility is the  use of several layers of different
 passive nuclear track detectors.  
\par
An exposure at a high altitude would effectively lower the monopole mass 
threshold 
and it would offer the new possibility of searching
for any heavily ionizing object present in the cosmic radiation
and which has 
a strong interaction cross section. This includes light MMs with
attached $p$ or nuclei, dyons,  
nuclearites and Q-balls; in fact one might consider this last point as one
of the main 
reasons for such a search at high altitude. \par

 The CR39 nuclear track detector allows to search for magnetic monopoles
 with one 
unit Dirac charge ($g_D$),
 for  
$\beta =v/c$ around $10^{-4}$   and for  $\beta >10^{-3}$, 
the whole  $\beta$-range of 
$4 \times 10^{-5}<\beta< 1$ for MMs with $g \geq 2 g_D$,
 for dyons, for nuclearites and for Q-balls.  
\par

We are presently making  
tests by exposing nuclear track detectors in Bologna and at 
the Chacaltaya mountain station, 
 in order to study the effects of possible backgrounds and of possible climatic 
conditions.

\vskip 1.0 truecm
 \noindent{\bf 3. Proposal}\par
\vskip 0.5 truecm

As already stated, we would like to implement passive nuclear
track detectors of 400 m$^2$  at mountain altitude.
  The track-etch detectors could be organised in modules of 24 cm $\times$ 24 
cm, each made of 3 layers of CR39, 3 layers 
of polycarbonate and of an 
aluminium absorber  1 mm thick; this module would be placed  in an
aluminized polyethilene bag filled with dry air.
These bags  reduce by about one order of magnitude the radon
background.
The CR39 is the main nuclear 
track detector; the polycarbonate has a higher threshold, and it is useful for high 
velocity monopoles and for nuclearites and 
 Q-balls with $\beta>10^{-4}$.\par

 The best 
and least expensive CR39 is produced by the
Intercast Europe Co. of Parma. We are 
in a position  to obtain very good material,   controlled continuously 
by us, and at the best possible 
price. \par

A program for the analysis of various types of polycarbonate (Lexan, Makrofol)
  is well under way.  So far the best results were obtained with Makrofol 
(made by Bayer) of 0.5 mm thickness.
There is no problem for the availability and cost either of Lexan or Makrofol.\par

  Once  exposed, CR39 and the Makrofol should be etched. 
Presently there are two etching facilities, one at Gran Sasso 
(the apparatus was built
in Torino) and one in Bologna. The 
etching capacity of the Gran Sasso apparatus is about 26 m$^2$/month for
``strong" etching of the first layer of CR39.
The Bologna apparatus is presently used for ``normal" etching of the 
second CR39 layers; we are planning to use it also
 for strong etching to complement the Gran Sasso apparatus. The global etching
 rate would be 42 m$^2$/month
At the quoted rate, we should be able to complete
the etching of the MACRO nuclear track detector in about 2 years. \par
For the SLIM experiment we plan to etch  small samples in the next few 
years, and start the main etching after the completion of the MACRO effort. 
 We propose to etch about
150 m$^2$/year by
using only the Bologna apparatus.
The collaborators of the Pinstech Lab. could take care of the 
exposure, etching and analysis
 of about 100 m$^2$ of detector.
\par
For the etching and for the analysis of the detector we may use 
the same methods presently used for MACRO; we shall also study other more
authomated methods.

 After exposure, the first sheet of each CR39 module is etched  using 
a ``strong etching" at 80$^{\circ}$C in an 8N water solution of NaOH.
 \par

  The etched CR39 is analyzed, first quickly using a 
simple method with a light source and a large lens and observing the  CR39 foil
  in  transparency; a more accurate analysis is later performed  
 with a large field binocular microscope.
 At present, for the MACRO CR39, which has an average exposure time of 
7 years, we find ``candidates'' in about 
8\% of the sheets. In these cases
we etch the second CR39 foil 
of the interested module at 70$^{\circ}$C in a 6N NaOH water 
solution (this is presently done in Bologna). 
\par
  We need to make regular small exposures of CR39 and Makrofol  
to relativistic heavy
ions in order to check the quality of the material used, its stability in time,
the absence of fading effects, etc. \cite{cali,cecco}.
One main calibration test  using Fe ions of 1 GeV/nucleon 
from the Brookhaven AGS is in progress.
\vskip 1.0 truecm
 \noindent{\bf 5. The collaboration }\par
\vskip 0.5 truecm

The present collaboration involves 
groups from INFN and the Universities of Bologna and  
 Torino in Italy, the University of Alberta in Canada, the 
University of Oujda in Morocco, the 
Pinstech Lab. in Islamabad, Pakistan,
 the Institute for Space Sciences of Bucharest, Romania, and
the 
Laboratorio de Fisica Cosmica de Chacaltaya in La Paz, Bolivia. 
The main contributions from the Bucharest and Oujda groups 
are with personnel often stationed
in Bologna in the context of bilateral agreements$^*$.

\vskip 1.0 truecm
 \noindent{\bf 7. Conclusions
}\par
\vskip 0.5 truecm
We propose to install a  nuclear track detector of
 400 m$^2$
 at the Chacaltaya high altitude lab. The detectors will 
be operated  for at least 4 years. They  will allow for a search 
 for light monopoles and dyons 
at the level of 
the Parker bound, that is at a flux of a $10^{-15}$ cm$^{-2}$ 
s$^{-1}$ sr$^{-1}$. A similar level of sensitivity will be obtained for 
relatively light nuclearites and for Q-balls. 

\vskip 1.0 truecm
 \noindent{\bf Acknowledgements.}
We would like to acknowledge fruitful discussions with A. De R\'{u}jula, S. 
Glashow, J. Steinberger, and other colleagues.
\par
\vskip 0.5truecm
{\footnotesize
$^*$We thank the University of Bologna, the ICTP-TRIL program, Worldlab and 
FAI-INFN for providing fellowships and grants for non Italian scientists.}


\begin{figure}
\begin{center}
        \mbox{ \epsfysize=15cm
            \epsffile{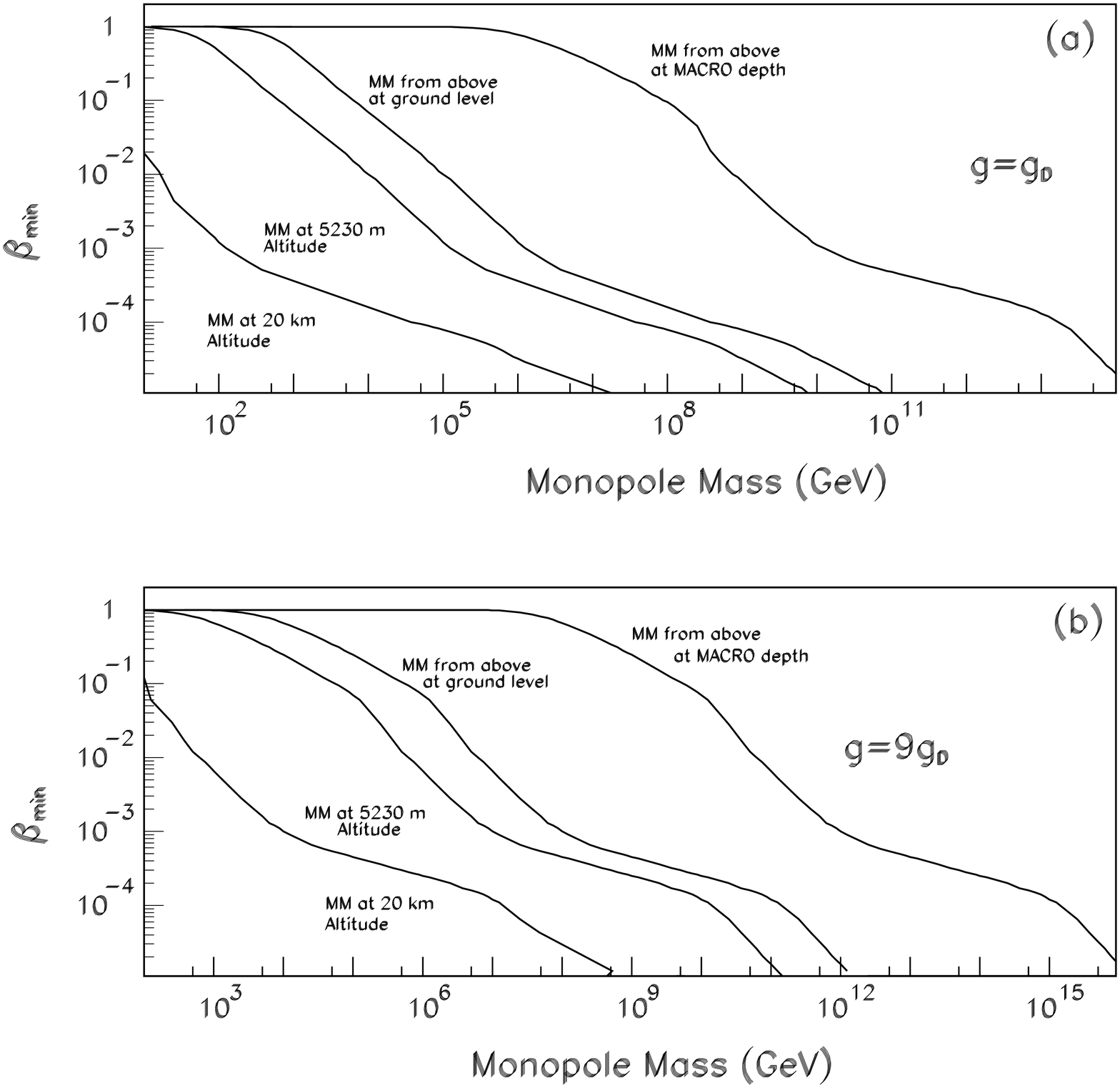}}

\end{center}
  \caption{\small Accessible regions in the plane (mass, $\beta$) for monopoles
with magnetic charge (a) $g=g_D$ and (b)  $g=~9 g_D$ coming from
above for an experiment at  altitudes of 20000 m, 5230 m, at sea 
level and for an underground
detector at the Gran Sasso Lab. (like the MACRO detector). It is assumed that
 monopoles interact only via the electromagnetic interaction, and no radiative
 effects are considered [9]. If a light monopole attaches a
nucleon or if it has some strong interaction, it could interact in the higher
atmosphere, like primary protons and nuclei of the cosmic radiation.}
  \label{fig:1}
\end{figure}

\begin{figure}
\begin{center}
        \mbox{ 
	\hspace{-1cm}
	\epsfysize=12.3cm
            \epsffile{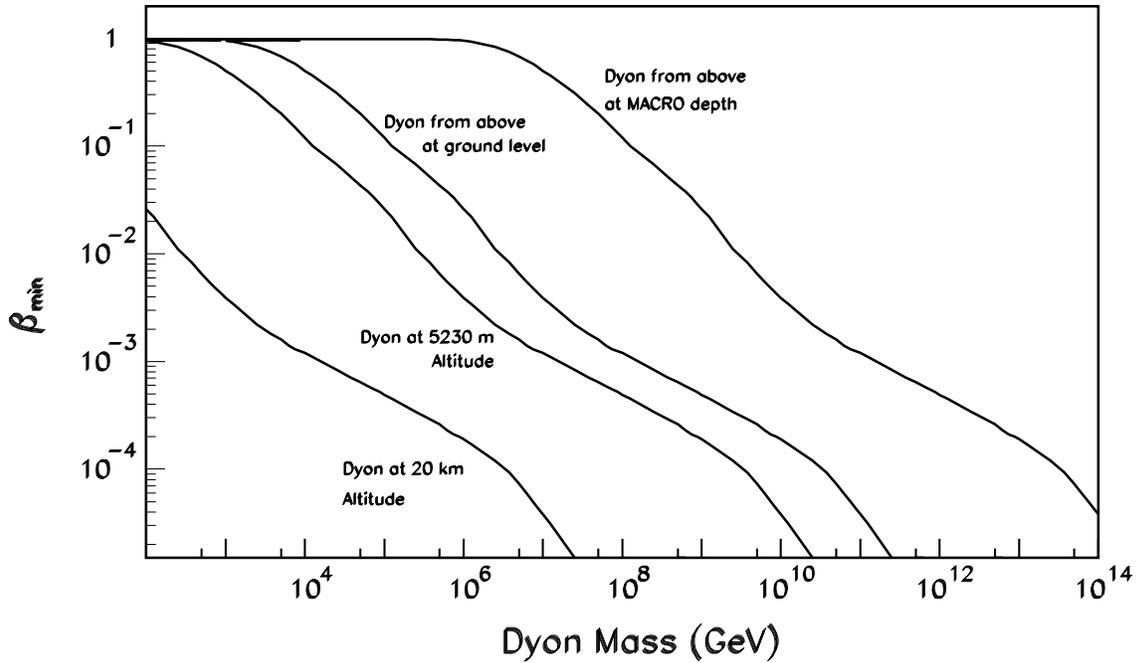}}
\end{center}
\vspace{-1cm}
   \caption{\small Accessible region in the plane (mass, $\beta$) for dyons 
from above for an experiment at  altitudes of 20000 m, 5230 m, at sea 
level and for an underground
detector at the Gran Sasso Lab. It is assumed that dyons interact only via
the electromagnetic interaction, and no radiative effects are considered [9].
 A monopole-proton composite would behave like a dyon with the addition of
a strong interaction cross section, which may make it to interact in the
higher atmosphere like primary protons and nuclei of the cosmic radiation.}
  \label{fig:2}
\end{figure}

\begin{figure}
\begin{center}
        \mbox{ \hspace{-1.5cm} \epsfysize=11cm
            \epsffile{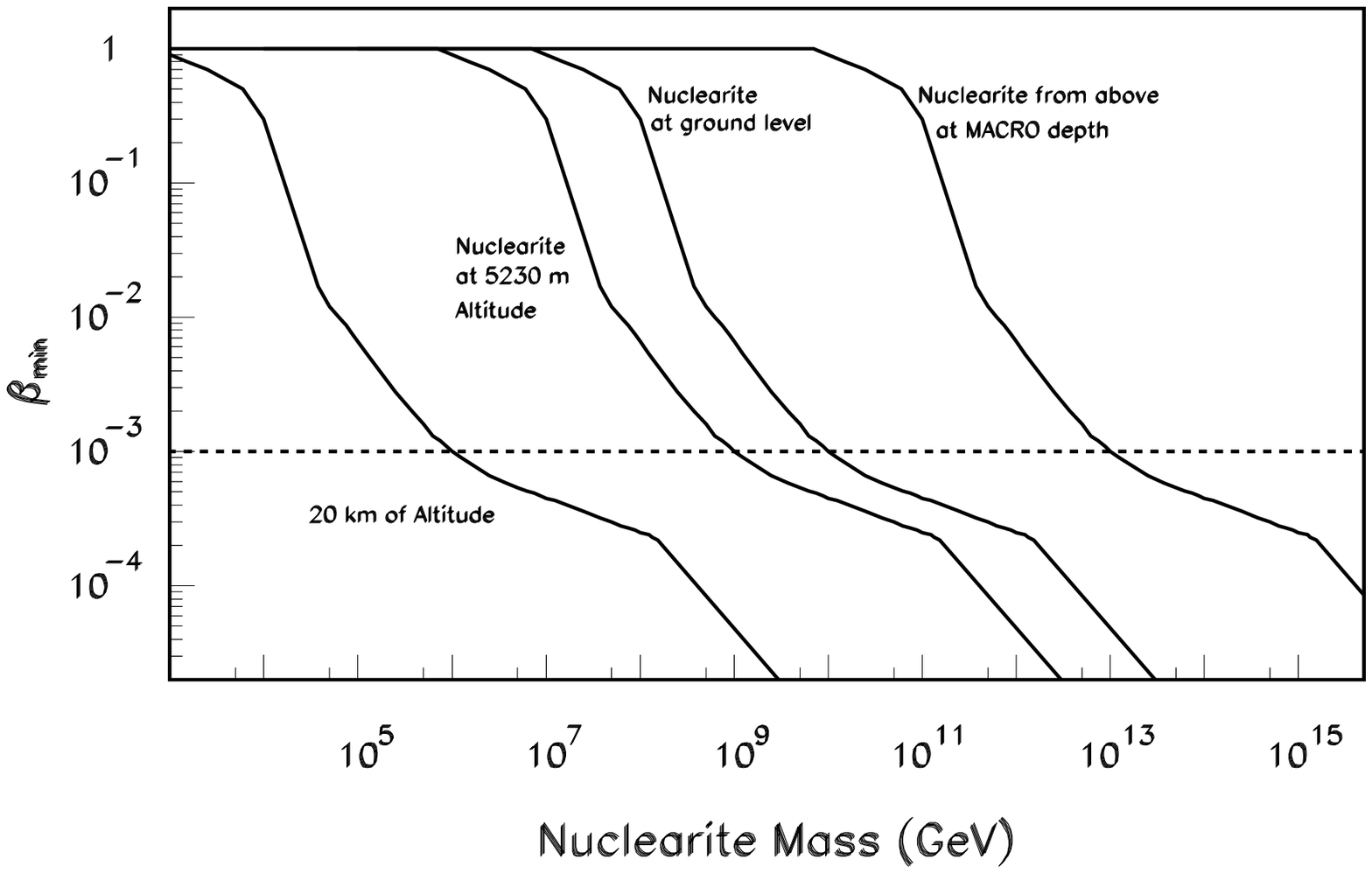}}
\end{center}
  \caption{\small Accessible region in the plane (mass, $\beta$) 
for nuclearites. Only the energy losses via electromagnetic interaction
are considered (bremsstrahlung is not considered). If nuclearites are part of
 the Dark Matter they would have typical velocities of $\beta \sim 10^{-3}$. 
 It is highly probable that low mass nuclearites interact strongly; thus they
may not reach the lower atmosphere.}
\end{figure}

\begin{thebibliography}{444}
\parskip -5pt
\bibitem{gg} Monopoles in Quantum Field Theory, N.S. Craige, P. Goddard, 
W. Nahm 
editors, World Scientific (1982); G. Giacomelli, Rivista del Nuovo Cimento 7  
(1984) 1.  Monopoles in Gauge Theories, Lecture Notes by N. Craige, 
G. Giacomelli, W. Nahm and Q. Shafi, World Scientific (1986).
\bibitem{macro} MACRO Collaboration (Search for magnetic monopoles with the 
MACRO detector) paper
presented at the 26$^{th}$ ICRC, Salt Lake City, Utah.
\bibitem{huguet} E. Huguet and P. Peter (Bound states in monopoles: 
sources for UHECR?) hep-ph/9901370 (1999).
\bibitem{weiler} T. J. Weiler and T. K. Kephart (Are we seeing magnetic
monopole cosmic rays at $E >~ \sim 10^{20}$ eV?), Nucl. Phys. B. Proc. Suppl.
51 (1996) 218; T. W. Kephart and T. J. Weiler (Magnetic monopoles
as the highest energy cosmic ray primaries), Astrop. Phys. 4 (1996) 217.
\bibitem{escobar} C.~O.~Escobar and R.~A.~Vazquez (Are high energy cosmic 
rays monopoles?), Astrop. Phys. 10 (1999) 197.
\bibitem{ruhula} A. 
De R\'{u}jula (Effects of virtual monopoles), Nucl. Phys. B435 (1995) 257.
\bibitem{parker} M. S. Turner, E. M. Parker and T. T. Bogdan (Magnetic 
monopoles and the survival of galactic magnetic 
fields), Phys. Rev. D26 (1982) 1296.
\bibitem{bps} P. M. Sutcliffe (BPS monopoles), Int. Journal of Modern Physics
  A12 (1997) 4663;
G. Lazarides et al. (Magnetic monopoles from superstring models), Phys. 
Rev. Lett. 58 (1987) 1707.
\bibitem{macron} M. Ambrosio et al. (Search for nuclearites with the MACRO 
detector at Gran Sasso), INFN/AE-97/20 (1997); in
Proc. of the 25$^{th}$ ICRC, Durban, vol. 7 (1997) 177.
\bibitem{wit} E. Witten, Phys. Rev. D30 (1984) 272; A. De R\'{u}jula and S. L.
 Glashow (Nuclearites-a novel form of cosmic radiation),
	Nature 312 (1984) 734.
\bibitem{qb} S. Coleman (Q-Balls), HUTP-85/A050 (1985); 
	A. Kusenko et al. (Experimental signatures of supersymmetric dark-matter
	Q-balls), CERN-TH/97-346, hep-ph/9712212 (1997).
\bibitem{derkaoui} J. Derkaoui et al. (Energy losses of magnetic monopoles and dyons in 
	the Earth), DFUB 98/1 (1998), Astrop. Phys. 9 (1998) 173.\\
	J. Derkaoui et al. (Energy losses of magnetic monopoles and dyons in 
	scintillators, streamer tubes and nuclear track detectors), DFUB 98/13, 
	submitted to Astrop. Phys. (1998).
\bibitem{naka} S. Nakamura et al. (A new limit on the flux of strange
matter), Phys. Lett. B263 (1991) 529.
\bibitem{wilk3}G. Wilk and Z. Wlodarczyk (Propagation of Strange Quark 
Matter in the Atmosphere), presented at Strangeness'96 Workshop, 
Budapest 1996, hep-ph/9606401 (1996).
\bibitem{centauro} C. M. G. Lattes et al. (Hadronic interactions of high-energy
cosmic ray observed by emulsion chambers), Phys. Rep.  65 (1980) 151;
A. S. Borisov et al. (Superhigh-energy cosmic ray interactions observed
in emulsion chambers at Pamir and mt. Chacaltaya), Phys. Lett. B190 (1987) 226.
\bibitem{wilk}G. Wilk and Z. Wlodarczyk (Centauro as Probe of Deeply 
Penetrating Component in Cosmic Rays), presented at IX$^{th}$ Int. 
Symp. on Very High Energy Cosmic Rays, Karlsruhe, 1996 and 
hep-ph/9609228 (1996)
\bibitem{wilk2}G. Wilk and Z. Wlodarczyk (How can strange quark matter 
occur deeply in the atmosphere?), hep-ph/9603228 (1996).
\bibitem{macrom} 
M. Ambrosio et al., MACRO Coll. (MACRO limits for $g=g_D$ magnetic monopoles), 
MA\-CRO/PUB 96/2 (1996); (Magnetic monopole search with the MACRO detector at 
Gran Sasso), Phys. Lett. B486 (1997) 249; (Magnetic monopole 
search with the MACRO detector at Gran Sasso), INFN/AE-97/19 (1997), in 
Proc. of the 25$^{th}$ ICRC, Durban, vol. 7 (1997) 157; F. Cei, for the MACRO 
Coll. (Search for rare particles with the MACRO detector), High Energy Physics 
Int. Conf., Vancouver, Canada (1998).
\bibitem{rep} G. Giacomelli, (MACRO Status Report),
 LNGS 1998 Annual Report;  (1998);
 M. Ambrosio et al., MACRO Coll.,  (Nuclearite
search with the MACRO detector at Gran Sasso) hep-ex/9904031, 
EPJ C (in press) (1999); see also
 Proc. of the 26$^{th}$ ICRC, Salt Lake City, vol.~2 (1999) 5; 112; 168; 
172; 213; 277; 332.
 



\bibitem{cali} G. Giacomelli et al. (Extended calibration of a CR39 
nuclear track detector with 158 AGeV Pb$^{207}$ ions), 
Nucl. Instr.  Meths. A411 (1998) 41.
\bibitem{cecco} S. Cecchini et al. 
 (Calibration with Relativistic
    and Low Velocity Ions of a CR39 Nuclear Track Detector), 
Nuovo Cimento 109 (1996) 1119.


\end{thebibliography}
\end{document}